\documentclass[conference]{IEEEtran}
\usepackage{graphicx}
\usepackage{epstopdf}
\usepackage{citesort}
\usepackage{cite}
\usepackage{amsmath}
\usepackage{algorithmic}

\usepackage{amsfonts}
\usepackage{epstopdf}
\usepackage{xcolor}
\usepackage{amssymb,bm,upgreek}
\usepackage{algorithm}
\usepackage{algorithmic}
\usepackage{multirow}
\usepackage{transparent}
\usepackage{verbatim}

\usepackage{import}

\DeclareMathOperator\erf{erf}

\newcommand{\RX}{\textnormal{RX}}
\newcommand{\R} {\textnormal{RX}}
\newcommand{\TX}{\textnormal{TX}}
\newcommand{\T}{\textnormal{TX}}
\newcommand{\FC} {\textnormal{FC}}
\newcommand{\trans}{\textrm{trans}}
\newcommand{\report}{\textrm{report}}
\newcommand{\metre}{\textnormal{m}}
\newcommand{\s}{\textnormal{s}}

\newcommand{\m}{\textnormal{m}}

\newcommand{\ob}{\textnormal{ob}}
\newcommand{\md}{\textnormal{md}}
\newcommand{\fa}{\textnormal{fa}}

\begin{document}
\title{Simplified Cooperative Detection for Multi-Receiver Molecular Communication}

\author{\IEEEauthorblockN{Yuting Fang${}^\dag$, Adam Noel${}^\ddag$, Yiran Wang${}^\dag$, Nan Yang${}^\dag$}

\IEEEauthorblockA{${}^\dag$Research School of Engineering, Australian National University, Canberra, ACT, Australia\\}
\IEEEauthorblockA{${}^\ddag$School of Electrical Engineering and Computer Science, University of Ottawa, Ottawa, ON, Canada\\}}
\IEEEspecialpapernotice{(Invited Paper)}
\maketitle

\maketitle

\begin{abstract}
Diffusion-based molecular communication (MC) systems experience significant reliability losses. To boost the reliability, a MC scheme where multiple receivers (RXs) work cooperatively to decide the signal of a transmitter (TX) by sending the same type of molecules to a fusion center (FC) is proposed in this paper. The FC observes the total number of molecules received and compares this number with a threshold to determine the TX's signal. The proposed scheme is more bio-realistic and requires relatively low computational complexity compared to existing cooperative schemes where the RXs send and the FC recognizes different types of molecules. Asymmetric and symmetric topologies are considered, and closed-form expressions are derived for the global error probability for both topologies. Results show that the trade-off for simplified computations leads to a slight reduction in error performance, compared to the existing cooperative schemes.
\end{abstract}
\section{Introduction}
Molecular communication (MC) describes how information is exchanged using molecules, which is one of the most common means of communication among biological entities \cite{nakano2013molecular}.
The unique features of MC enable it to advance nano-applications in a variety of fields, such as the analysis of biological materials \cite{52}, the engineering of tissue structure \cite{55}, the interface between human brain and electrical devices \cite{56}, and the targeted drug delivery \cite{57}.

Researchers are developing theoretical models to analyze and improve the quality of MC systems. Of all the existing theoretical models, the diffusion-based model is a simple and fundamental one. In such a system, the communication depends only on the random walk of the information molecules; there is no additional mechanism required. However, a common issue is that the reliability of a single-link MC system rapidly decreases when the propagation distance increases. To solve this problem, one approach where multiple receivers (RXs) share common information, was proposed and is commonly found in biology. For example, nitric oxide is a gas particle that passes from its source to neighboring cells, binds to the receptor guanylyl cyclase of neighboring cells, and activates the receptors to perform the synthesis of messenger molecules \cite[Ch~7]{raven2008biology}. Another example is the protein Interleukin-6, which is detectable by many types of cells such as B cells and T cells. The protein supports these cells and thus enables functions in the immune system \cite{kishimoto1243}.

The majority of existing MC studies have focused on the modeling of single-link MC systems. To solve the reliability issue, some papers, such as \cite{7485823,atakan2008molecular,7397863}, have investigated multi-hop or multi-RX MC systems that build upon single links.
\cite{7485823} and \cite{atakan2008molecular} considered multi-transmitter (TX) networks. In \cite{7397863}, various techniques for a multi-input multi-output MC system were proposed. However, the benefits of cooperation among multiple RXs to determine a TX's symbol sequence have not been studied. An exception is our work described in \cite{fang2016distributed,fang2016convex,fang2017maximum}. \cite{fang2016distributed} and \cite{fang2016convex} analyzed the error performance of a cooperative MC system where multiple RXs report their decisions on a TX's symbols to a fusion center (FC) using distinct types of molecules. The FC uses hard fusion rules to make a final decision. \cite{fang2017maximum} considered maximum likelihood detection in a multi-RX systems to determine the lower bounds on the error performance that can be achieved using the hard fusion rules considered in \cite{fang2016distributed} and \cite{fang2016convex}. However, multiple types of molecules may not be available in some biological environments. Also, identifying different types of molecules and performing maximum likelihood detection at the FC may be cumbersome in certain applications.

In this paper, we consider a cooperative MC system where multiple RXs report their decisions on a TX's symbols to a FC using the \emph{same} type of molecule and the FC makes a global decision by comparing the observations with a constant threshold. For the sake of convenience, we call this cooperative scheme decode and forward (DF) with single-molecule-type and constant threshold at the FC (SD-Constant). Our goal is to demonstrate the reliability improvement over a single link even though we have a constraint on the types of molecules available at the RXs. Compared with the hard fusion rules considered in \cite{fang2016distributed}, the SD-Constant scheme is more suitable for the environment where the processing capabilities of devices are more limited, or where the number of types of molecules available is constrained. We consider asymmetric and symmetric topologies for the cooperative MC system. For both topologies, we derive closed-form analytical expressions for the analytical global error probabilities. Using numerical and simulation results, we validate our analytical results and show that the error performance of the SD-Constant scheme is better than that of a single link.
\begin{figure}
  \centering
  \includegraphics[height=1.5in]{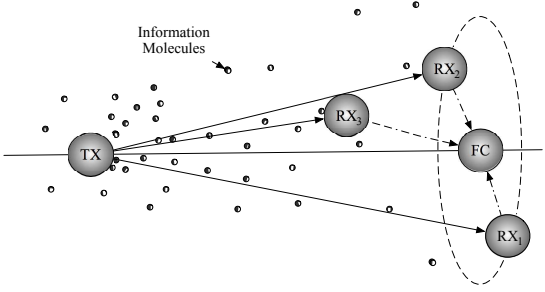}
  \caption{Illustration of the considered cooperative MC system with $K = 3$ for the asymmetric topology. Solid arrows show the TX-RX links and dotted arrows show the RX-FC links. The direction of the arrows demonstrate the information flow. To ensure the asymmetric topology, the distance of link $\TX$-$\RX_3$ is different from those of the other two TX-RX links.}
  \label{systemmodel}
\end{figure}

\section{System Model}
We consider a three-dimensional diffusive MC system based on \cite{fang2016distributed}, with a point TX, a set of $K$ RXs, and an FC, as shown in Fig. \ref{systemmodel}. We generally assume that the RXs are closer to the FC than to the TX. We denote $d_{\scriptscriptstyle\T_k}$ and $d_{\scriptscriptstyle\FC_k}$ as the distance between the $k$th RX and the TX, and the distance between the $k$th RX and the FC, respectively. The RXs and the FC are passive spherical observers such that the information molecules diffuse through them without reacting. Accordingly, we denote $V_{\scriptscriptstyle\R_k}$ and $r_{\scriptscriptstyle\R_k}$ as the volume and the radius of the $k$th RX, $\RX_k$, respectively, where $k\in\{1,2,\ldots,K\}$. $V_{\scriptscriptstyle\FC}$ and $r_{\scriptscriptstyle\FC}$ are the volume and radius of the FC. We further assume that all individual observations are independent of each other. During each interval, the TX first releases type $A$ molecules, and then the RXs detect the number of the type $A$ molecules that are within the spheres. After that, all of the RXs report decisions using type $B$ molecules to the FC and the FC makes a final decision based on all information provided by the RXs. These are the three phases of the communication in our system, detailed as follows:

In the first phase, the TX transmits one symbol of information via the type $A$ molecules to the RXs through the diffusive channel. The number of released type $A$ molecules is denoted by $S_A$. We assume that each molecule diffuses independently. The TX uses ON/OFF keying to convey information, which means that the TX releases $S_{A}$ molecules of type $A$ to convey information symbol ``1'', and releases nothing to convey information symbol ``0''. The information transmitted by the TX is encoded into an $L$-length binary sequence, denoted by $\textbf{W}_{\scriptscriptstyle\T}=\{W_{\scriptscriptstyle\T}[1],W_{\scriptscriptstyle\T}[2],\ldots,W_{\scriptscriptstyle\T}[L]\}$, where $W_{\scriptscriptstyle\T}[j]$, $j\in\{1,\ldots,L\}$, is the $j$th symbol transmitted by the TX. We also assume that $\textrm{Pr}(W_{\scriptscriptstyle\T}[j]=1)=P_1$ and $\textrm{Pr}(W_{\scriptscriptstyle\T}[j]=0)=1-P_{1}$, where $\textrm{Pr}(\cdot)$ denotes probability. Once released, the type $A$ molecules diffuse freely in the environment and are detectable by all RXs.

In the second phase, each RX makes binary decisions (hard decisions) of ``0'' and ``1'' by comparing the number of observed molecules with a constant threshold. The decision of the $k$th RX on the $j$th symbol is ${\hat{W}_{\scriptscriptstyle\RX_k}}[j]$. Once the decisions are made, all $K$ RXs simultaneously release type $B$ molecules to report to the FC. We assume that the type $B$ molecules are released from the centers of the RXs. Similar to the TX, each RX uses ON/OFF keying to report its decision, i.e., the RX releases $S_B$ molecules of type $B$ if ${\hat{W}_{\scriptscriptstyle\RX_k}}[j]=1$, otherwise the RX releases no molecules. The released type $B$ molecules are only detectable by the FC.

In the final phase, the FC detects and counts all of the type $B$ molecules sent from RXs. The FC is not able to differentiate which molecules were released by which RXs. Thus, it compares the total number with a constant threshold to determine the current symbol transmitted by the TX, $\hat{W}_{\scriptscriptstyle\FC}[j]$. We define $\textbf{W}_{{\scriptscriptstyle\T}}^{l}=\{W_{\scriptscriptstyle\T}[1],\ldots,W_{\scriptscriptstyle\T}[l]\}$ as an $l$-length subsequence of the information transmitted by the TX, where $l\leq{L}$. We also define $\hat{\textbf{W}}_{\scriptscriptstyle\RX_k}^l=\{\hat{W}_{\scriptscriptstyle\RX_k}[1],\ldots,\hat{W}_{\scriptscriptstyle\RX_k}[l]\}$ as an $l$-length subsequence of the local hard decisions at $\RX_k$ and $\hat{\textbf{W}}_{\scriptscriptstyle\FC}^{l}=\{\hat{W}_{\scriptscriptstyle\FC}[1],\ldots,\hat{W}_{\scriptscriptstyle\FC}[l]\}$ as an $l$-length subsequence of the global decisions at the FC.


We synchronize the devices in the following way. The time interval for transmitting adjacent symbols is $T$. 
The RXs operate in half-duplex mode; they take $M_{\scriptscriptstyle\RX}$ samples between $(j-1)T$ and $(j-1)T+t_{\text{trans}}$, where $t_{\trans}$ is the transmission time interval from the TX to the RXs, and report at times $(j-1)T+t_{\text{trans}}$, $m \in \{1,2,...,M_{\scriptscriptstyle\RX}\}$. The samples at the RXs are spaced equally, such that the $m$th RX sample for the $j$th symbol is taken at $t_{\scriptscriptstyle\RX}(j,m) = (j-1)T+m\Delta t_{\scriptscriptstyle\RX}$ where $\Delta{t_{\scriptscriptstyle\RX}}$ is the duration between two adjacent samples. We denote $t_{\text{report}}$ as the reporting time interval from the RXs to the FC, and we take the FC's $\tilde m$ samples at times $t_{\scriptscriptstyle\FC}(j,\tilde {m}) = (j-1)T+t_{\text{trans}}+\tilde{m}\Delta t_{\scriptscriptstyle\FC}$, $\Delta t_{\scriptscriptstyle\FC}$ being the time duration between two adjacent FC samples and $\tilde m \in \{1,2,...,M_{\scriptscriptstyle\FC}\}$. We assume $M_{\scriptscriptstyle\RX}\Delta t_{\scriptscriptstyle\RX}<t_{\text{trans}}$ to ensure the half-duplex mode. We also assume that $M_{\scriptscriptstyle\FC}\Delta t_{\scriptscriptstyle\FC}<t_{\text{report}}$. We further assume that samples are combined using energy detection to reduce the computational complexity.

\section{Error Performance Analysis of Cooperative MC Systems}
In this section, we first review the analytical error performance of the $\TX$-$\RX_k$ link 
and then analyze the analytical global error performance of the cooperative MC system for asymmetric and symmetric topologies. In the asymmetric topology, the $d_{\scriptscriptstyle\TX_k}$ are not identical for all $\RX_k$ and/or the $d_{\scriptscriptstyle\FC_k}$ are not identical for all $\RX_k$. In the symmetric topology, the $d_{\scriptscriptstyle\TX_k}$ are identical for all $\RX_k$, and the $d_{\scriptscriptstyle\FC_k}$ are also identical for all $\RX_k$.

\subsection{$\TX$-$\R_k$ $\textnormal{Link}$}
In this subsection we examine the analytical error performance of the $\TX$-$\RX_k$ link, based on the analytical methods presented in \cite{6868273}. We first evaluate $P_{\ob}^{({\scriptscriptstyle{\T},{\scriptscriptstyle\R_k}})}\left(t\right)$,  the probability of observing a type $A$ molecule inside $V_{\scriptscriptstyle\R_k}$ at time $t$, where the molecule was emitted from the TX at $t=0$. Assuming that type $A$ molecules diffuse freely from a sufficient distance TX to the RX, based on \cite{6868273}, we evaluate $P_{\ob}^{({\scriptscriptstyle{\T},{\scriptscriptstyle\R_k}})}\left(t\right)$ as
\begin{equation}\label{simplediffusion}
P^{(\scriptscriptstyle\TX,\scriptscriptstyle\RX_{k})}_{\ob} (t)= \frac{V_{\scriptscriptstyle\RX_k}}{(4\pi D_At)^{3/2}}\exp\left(-\frac{d^2_{\scriptscriptstyle\TX_k}}{4D_At}\right),
\end{equation}
where $D_A$ is the diffusion coefficient of the type $A$ molecules and $d_{\scriptscriptstyle\TX_k}$ is the distance between the TX and the center of $\RX_k$. The number of molecules observed within $V_{\scriptscriptstyle\R_k}$ in the $j$th symbol interval due to the emission of molecules from the current and previous symbol intervals at the TX, $\textbf{W}_{\scriptscriptstyle\T}^j$, is $S_{\text{ob}}^{(\scriptscriptstyle\TX,\scriptscriptstyle\RX_k)}[j]$. As per \cite{6868273}, $S_{\text{ob}}^{(\scriptscriptstyle\TX,\scriptscriptstyle\RX_k)}[j]$ is approximated by a Poisson random variable (RV) where the mean is
\begin{align}\label{observed molecular numbers R}
\bar{S}_{\text{ob}}^{(\scriptscriptstyle\TX,\scriptscriptstyle\RX_k)}[j]
= S_A\sum\limits^{j}_{i=1}W_{\scriptscriptstyle\TX}[i]\sum\limits^{M_{\scriptscriptstyle\RX}}_{m=1}P_{\text{ob}}^{(\scriptscriptstyle\TX,\scriptscriptstyle\RX_k)}((j-i)T + m\Delta{t_{\scriptscriptstyle\RX}}).
\end{align}
Then, the $\text{RX}_k$ decides on the $j$th symbol by
\begin{equation}
\hat{W}_{\scriptscriptstyle\text{RX}_k}[j] =
  \begin{cases}
    1       & \quad \text{if } {S}^{(\scriptscriptstyle\TX,\scriptscriptstyle\RX_k)}_{\text{ob}}[j]\geq \xi_{\scriptscriptstyle\RX_k},\\
    0  & \quad \text{otherwise, }\\
  \end{cases}
\end{equation}
where $\xi_{\scriptscriptstyle\RX_k}$ is the constant threshold at $\R_k$, and is independent of the symbol intervals. Afterwards, we derive the analytical miss detection probability $P_{\text{md},k}[j]$ and false alarm probability $P_{\text{fa},k}[j]$ of the $\TX$-$\R_k$ link in the $j$th symbol interval, for a \emph{given} transmitter sequence\footnote{All the expected error probabilities throughout this paper are derived for given $\textbf{W}_{\scriptscriptstyle\T}^{j-1}$, unless otherwise specified.} $\textbf{W}_{\scriptscriptstyle\T}^{j-1}$. The case that ``1'' is transmitted but ``0'' is received is a miss detection and the case that ``0'' is transmitted but ``1'' is received is a false alarm. Based on \cite{6868273}, $P_{\text{md},k}[j]$ and $P_{\text{fa},k}[j]$ are written as
\begin{equation}
	P_{\text{md},k}[j] = \Pr\left(S_{\text{ob}}^{(\scriptscriptstyle{\TX,\RX}_k)}[j]<\xi_{\scriptscriptstyle\RX_k} |W_{\scriptscriptstyle\TX}[j]=1,\textbf{W}_{\scriptscriptstyle\TX}^{j-1}\right)
\end{equation}
and
\begin{equation}
	P_{\text{fa},k}[j] = \Pr\left(S_{\text{ob}}^{(\scriptscriptstyle{\TX,\RX}_k)}[j]\geq\xi_{\scriptscriptstyle\RX_k} |W_{\scriptscriptstyle\TX}[j]=0,\textbf{W}_{\scriptscriptstyle\TX}^{j-1}\right).
\end{equation}

\subsection{Global Error Probability}
In this subsection, we determine the global error performance of the cooperative MC system with the SD-Constant scheme for asymmetric and symmetric topologies. The analytical global error probability in the $j$th symbol interval for a \emph{given} TX sequence $\textbf{W}_{\scriptscriptstyle\T}^{j-1}$, $Q_{\scriptscriptstyle\FC}[j]$, is
\begin{equation}\label{globalerrorprobability}
Q_{\scriptscriptstyle\FC}[j] = P_1Q_{\text{md}}[j] + (1-P_{1})Q_{\text{fa}}[j],
\end{equation}
where $Q_{\text{md}}[j]$ and $Q_{\text{fa}}[j]$ are the analytical global miss detection and the analytical global false alarm probabilities in the $j$th symbol interval, respectively. By averaging $Q_{\scriptscriptstyle\FC}[j]$ over the number of all possible realizations of $\textbf{W}_{\scriptscriptstyle\T}^{j-1}$ and all symbol intervals, the analytical average error probability of the cooperative MC system, $\overline{Q}_{\scriptscriptstyle\FC}$, is obtained.

We denote $P_{\ob}^{({\scriptscriptstyle\R_k,\scriptscriptstyle\FC})}(t)$ as the probability of observing a given $B$ molecule, emitted from the center of $\RX_k$ at $t=0$, inside $V_{\scriptscriptstyle\FC}$ at time $t$. Due to the close distance between the RXs and the FC, the uniform concentration assumption of the $B$ molecules inside $V_{\scriptscriptstyle\FC}$ is not valid. Thus, we use \cite[Eq. (27)]{Adamdimen} to evaluate $P_{\ob}^{(\scriptscriptstyle\R_k,\scriptscriptstyle\FC)}(t)$ as
\begin{align}\label{RXtoFCdiffusion}
P_{\ob}^{({\scriptscriptstyle\R_k,\FC})}(t) = &\;\frac{1}{2}\left[\erf\left(\frac{r_{\scriptscriptstyle\FC}+d_{\scriptscriptstyle\FC_k}}{2\sqrt{D_{{B}}t}}\right)+\erf\left(\frac{r_{\scriptscriptstyle\FC}-d_{\scriptscriptstyle\FC_k}}{2\sqrt{D_{B}t}}\right)\right]\nonumber\\
&-\frac{\sqrt{D_{B}t}}{d_{\scriptscriptstyle\FC_k}\sqrt{\pi}}\left[\exp\left(-\frac{(-d_{\scriptscriptstyle\FC_k}+r_{\scriptscriptstyle\FC})^{2}}{4D_{{B}}t}\right)\right.\nonumber\\
&\left.-\exp\left(-\frac{(-d_{\scriptscriptstyle\FC_k}-r_{\scriptscriptstyle\FC})^{2}}{4D_{{B}}t}\right)\right],
\end{align}
where $D_B$ is the diffusion coefficient of the type $B$ molecules and $d_{\scriptscriptstyle\FC_k}$ is the distance from $\RX_k$ to the FC. We then denote $S_{\text{ob}}^{(\scriptscriptstyle\RX_k,\scriptscriptstyle\FC)}[j]$ as the number of molecules that are observed by the FC in the $j$th symbol interval, due to the emission of molecules from the current and previous symbol intervals at the $\RX_k$, $\textbf{W}_{\scriptscriptstyle{\R_k}}^j$. Similar to $S_{\text{ob}}^{(\scriptscriptstyle\TX,\scriptscriptstyle\RX_k)}[j]$, ${S}_{\ob}^{({\scriptscriptstyle\R_k,\scriptscriptstyle\FC})}[j]$ can also be accurately approximated by a Poisson RV. We denote ${\bar{S}}_{\ob}^{({\scriptscriptstyle\R_k,\scriptscriptstyle\FC})}[j]$ as the mean of ${S}_{\ob}^{({\scriptscriptstyle\R_k,\scriptscriptstyle\FC})}[j]$ and obtain it by replacing $S_{A}$, $W_{\scriptscriptstyle\T}[i]$, $P_{\ob}^{({\scriptscriptstyle{\T},{\scriptscriptstyle\R_k}})}$, $M_{\scriptscriptstyle\RX}$, $m$, and $\Delta{t_{\scriptscriptstyle\R}}$ with $S_{B}$, $\hat{W}_{\scriptscriptstyle\R_k}[i]$, $P_{\ob}^{({\scriptscriptstyle\R_k,\scriptscriptstyle\FC})}$, $M_{\scriptscriptstyle\FC}$, $\tilde{m}$, and $\Delta{t_{\scriptscriptstyle\FC}}$ in \eqref{observed molecular numbers R}.

We now derive $Q_{\md}[j]$ and $Q_{\fa}[j]$ for the asymmetric topology. We first define a set of decisions at all RXs in the $j$th symbol interval as $\hat{\textbf{W}}^{\scriptscriptstyle\RX}_j = \{\hat{W}_{\scriptscriptstyle\RX_1}[j],\hat{W}_{\scriptscriptstyle\RX_2}[j],...,\hat{W}_{\scriptscriptstyle\RX_K}[j]\}$. We then define a set $\mathcal{R}$ which includes all possible realizations of $\hat{W}^{\scriptscriptstyle\RX}_j$ and the cardinality of $\mathcal{R}$ is $2^K$. $Q_{\text{md}}[j]$ is the summation of the probabilities of miss detection associated with each realization of $\hat{W}^{\scriptscriptstyle\RX}_j$, scaled by the likelihood of that realization. Using this knowledge, we derive $Q_{\text{md}}[j]$ as
\begin{align}\label{globalmdasym}
Q_{\text{md}}[j] =&  \sum_{\hat{\textbf{W}}^{\scriptscriptstyle\RX}_j\in \mathcal{R}} \Bigg(\Pr\Big(\hat{\textbf{W}}^{\scriptscriptstyle\RX}_j\Big|W_{
\scriptscriptstyle\TX}[j]=1,{\textbf{W}^{j-1}_{\scriptscriptstyle\TX}}\Big)
\nonumber \times \\ & \Pr\Bigg(\sum^{K}_{k=1}S_{\text{ob}}^{(\scriptscriptstyle\RX_k,\scriptscriptstyle\FC)}[j]<\xi_{\scriptscriptstyle\FC}\Bigg|\hat{\textbf{W}}^{\scriptscriptstyle\RX}_j
,\hat{\textbf{W}}_{\scriptscriptstyle\RX_1}^{j-1},\nonumber\\& \hat{\textbf{W}}_{\scriptscriptstyle\RX_2}^{j-1},...,\hat{\textbf{W}}_
{\scriptscriptstyle\RX_K}^{j-1}\Bigg)\Bigg),
\end{align}
where $\xi_{\scriptscriptstyle\FC}$ is the constant threshold at the FC and is independent of symbol intervals. Similarly, \eqref{globalmdasym}, $Q_{\text{fa}}[j]$ is
\begin{align}\label{globalfaasym}
Q_{\text{fa}}[j] =&  \sum_{\hat{\textbf{W}}^{\scriptscriptstyle\RX}_j\in \mathcal{R}} \Bigg(\Pr\Big(\hat{\textbf{W}}^{\scriptscriptstyle\RX}_j\Big|W_{
\scriptscriptstyle\TX}[j]=0,{\textbf{W}^{j-1}_{\scriptscriptstyle\TX}}\Big)
\nonumber \times \\ & \Pr\Bigg(\sum^{K}_{k=1}S_{\text{ob}}^{(\scriptscriptstyle\RX_k,\scriptscriptstyle\FC)}[j]\geq\xi_{\scriptscriptstyle\FC}\Bigg|\hat{\textbf{W}}^{\scriptscriptstyle\RX}_j
,\hat{\textbf{W}}_{\scriptscriptstyle\RX_1}^{j-1},\nonumber\\& \hat{\textbf{W}}_{\scriptscriptstyle\RX_2}^{j-1},...,\hat{\textbf{W}}_
{\scriptscriptstyle\RX_K}^{j-1}\Bigg)\Bigg).
\end{align}

We now focus on the symmetric topology.
We define a subset $\mathcal{R}_n$, $n\in\{0,1,2,...,K\}$, which contains all of the realizations where $n$ RXs decide ``1'' and $K-n$ RXs decide ``0''. Since the distances of the $\TX$-$\RX_k$ links are identical in the symmetric topology, we write $P_{\text{md},k}[j] = P_{\text{md}}[j]$ and $P_{\text{fa},k}[j] = P_{\text{fa}}[j]$. We note that the distances of $\RX_k$-$\FC$ links are also identical in the symmetric topology. Using these notations, we then simplify \eqref{globalmdasym} and \eqref{globalfaasym} as
\begin{align}\label{global miss detection component}
Q_{\text{md}}[j] =&  \sum_{n=0}^{K}\Bigg[\binom Kn P_{\text{md}}[j]^{K-n}(1-P_{\text{md}}[j])^{n}
\nonumber \\&\times  \Pr\Bigg(\sum^{K}_{k=1}S_{\text{ob}}^{(\scriptscriptstyle\RX_k,\scriptscriptstyle\FC)}[j]<\xi_{\scriptscriptstyle\FC}\Bigg| \hat{\textbf{W}}^{\scriptscriptstyle\RX}_j\in \mathcal{R}_n
,\hat{\textbf{W}}_{\scriptscriptstyle\RX_1}^{j-1},\nonumber\\
& \hat{\textbf{W}}_{\scriptscriptstyle\RX_2}^{j-1},...,\hat{\textbf{W}}_
{\scriptscriptstyle\RX_k}^{j-1}\Bigg)\Bigg],
\end{align}
and
\begin{align}\label{global false alarm component}
    Q_{\text{fa}}[j] =&  \sum_{n=0}^{K}\Bigg[\binom Kn P_{\text{fa}}[j]^{K-n}(1-P_{\text{fa}}[j])^{n}\
    \nonumber \\&\times \Pr\Bigg(\sum^{K}_{k=1}S_{\text{ob}}^{(\scriptscriptstyle\RX_k,\scriptscriptstyle\FC)}[j]\geq\xi_{\scriptscriptstyle\FC}\Bigg| \hat{\textbf{W}}^{\scriptscriptstyle\RX}_j\in \mathcal{R}_n
,\hat{\textbf{W}}_{\scriptscriptstyle\RX_1}^{j-1},\nonumber\\
& \hat{\textbf{W}}_{\scriptscriptstyle\RX_2}^{j-1},...,\hat{\textbf{W}}_
{\scriptscriptstyle\RX_k}^{j-1}\Bigg)\Bigg].
\end{align}

\section{Numerical results and simulations}
In this section, we present our numerical results for the error performance of the proposed SD-Constant scheme, where the simulation results are generated by a particle-based simulation method. We set the simulation parameters as follows: the TX releases 8000 molecules for symbol ``1'' and each RX releases $\lceil2000/K\rceil$ molecules for decision ``1''
, i.e, the total number of molecules released by all RXs for symbol ``1'' is fixed at $2000$. We place the TX at ($0$, $0$, $0$) and the FC at ($2\,{\mu}\metre$, $0$, $0$). We also set up the locations of the RXs for the symmetric topology and the asymmetric topology as presented in Tables \ref{tab:coordinates1} and \ref{tab:coordinates2}.
Throughout this section, we only vary the threshold at the $\R_k$, $\xi_{\scriptscriptstyle\R_k}$, the threshold at the FC, $\xi_{\scriptscriptstyle\FC}$, the number of RXs, $K$, and the distance between the TX and the center of $\RX_k$, $d_{\scriptscriptstyle\TX_k}$. Other fixed parameters are listed in Table \ref{tab:table1}. Furthermore, we consider the same radius and the same detection threshold for all RXs such that $r_{\scriptscriptstyle\R_k}=r_{\scriptscriptstyle\R}$ and $\xi_{\scriptscriptstyle\RX_k} = \xi_{\scriptscriptstyle\RX}, \forall k$.
\begin{table}[t]
\renewcommand{\arraystretch}{1.2}
\centering
\caption{RXs' Location for the Symmetric Topology}\label{tab:coordinates1}
\begin{tabular}{l||c|c|c}
\hline
$\textbf{RXs}$ & $\textbf{X-axis}$ [${\mu}\metre$]& $\textbf{Y-axis}$ [${\mu}\metre$] & $\textbf{Z-axis}$ [${\mu}\metre$]\\\hline
\hline
$\RX_1$ & 2 & 0.6 & 0\\\hline
$\RX_2$ & 2 & -0.3& 0.5196 \\\hline
$\RX_3$ & 2 &  -0.3& -0.5196 \\\hline
$\RX_4$ & 2 & -0.6& 0 \\\hline
$\RX_5$ & 2 & 0.3 & 0.5196 \\\hline
$\RX_6$ & 2 & 0.3 & -0.5196 \\
\hline
\end{tabular}
\end{table}

\begin{table}[t]
\caption{Rxs' Location for the Asymmetric Topology}\label{tab:coordinates2}
\renewcommand{\arraystretch}{1.2}
\centering
\begin{tabular}{c||c|c|c}
\hline
$\textbf{RXs}$ & $\textbf{X-axis}$ [${\mu}\metre$]& $\textbf{Y-axis}$ [${\mu}\metre$] & $\textbf{Z-axis}$ [${\mu}\metre$]\\\hline
\hline
$\RX_1$ (fixed) & 2 & 0 & 0.6\\\hline
$\RX_2$ (fixed)& 2 & 0& -0.6\\\hline
$\RX_2$ (1)& 2 & 6& 0 \\\hline
$\RX_3$ (2)&1.6  & 0.48 & 0 \\\hline
$\RX_3$ (3)&1.2  & 0.36 & 0 \\\hline
$\RX_3$ (4)&0.8  & 0.24 & 0 \\\hline
$\RX_3$ (5)&0.4  & 0.12 & 0 \\\hline
\end{tabular}
\end{table}
\begin{table}[t]
\centering
\caption{Environmental Parameters Used in Simulations}\label{tab:table1}
\renewcommand{\arraystretch}{1.2}
\begin{tabular}{l||c|c}
\hline
$\textbf{Parameter}$ & $\textbf{Symbol}$& $\textbf{Value}$ \\\hline
\hline
Radius of RXs& $r_{\scriptscriptstyle\R}$ & $\{0.225\,{\mu}\metre,0.2\,{\mu}\metre\}$\\\hline
Radius of FC & $r_{\scriptscriptstyle\FC}$ & $0.225\,{\mu}\metre$ \\\hline
Time step at RX & $\Delta{t_{\scriptscriptstyle\R}}$ & $100\,{\mu}\s$\\\hline
Time step at FC & $\Delta{t_{\scriptscriptstyle\FC}}$ & $30\,{\mu}\s$ \\\hline
Number of samples of RX& $M_{\RX}$ & 5 \\\hline
Number of samples of FC& $M_{\FC}$ & 5 \\\hline
Transmission time interval & $t_{\trans}$ & $1\,{\m}\s$\\\hline
Report time interval & $t_{\report}$ & $0.3\,{\m}\s$\\\hline
Symbol interval time& $T$ & $1.1\,{\m}\s$\\\hline
Diffusion coefficient & $D_A=D_{B}$ & $5\times10^{-9}{\m^{2}}/{\s}$\\\hline
Length of transmitter sequence & $L$ & $10$ \\\hline
Probability of binary 1 & $P_1$ & $0.5$ \\
\hline
\end{tabular}
\end{table}
\begin{figure}[t]
  \centering
  \includegraphics[height=2.2in]{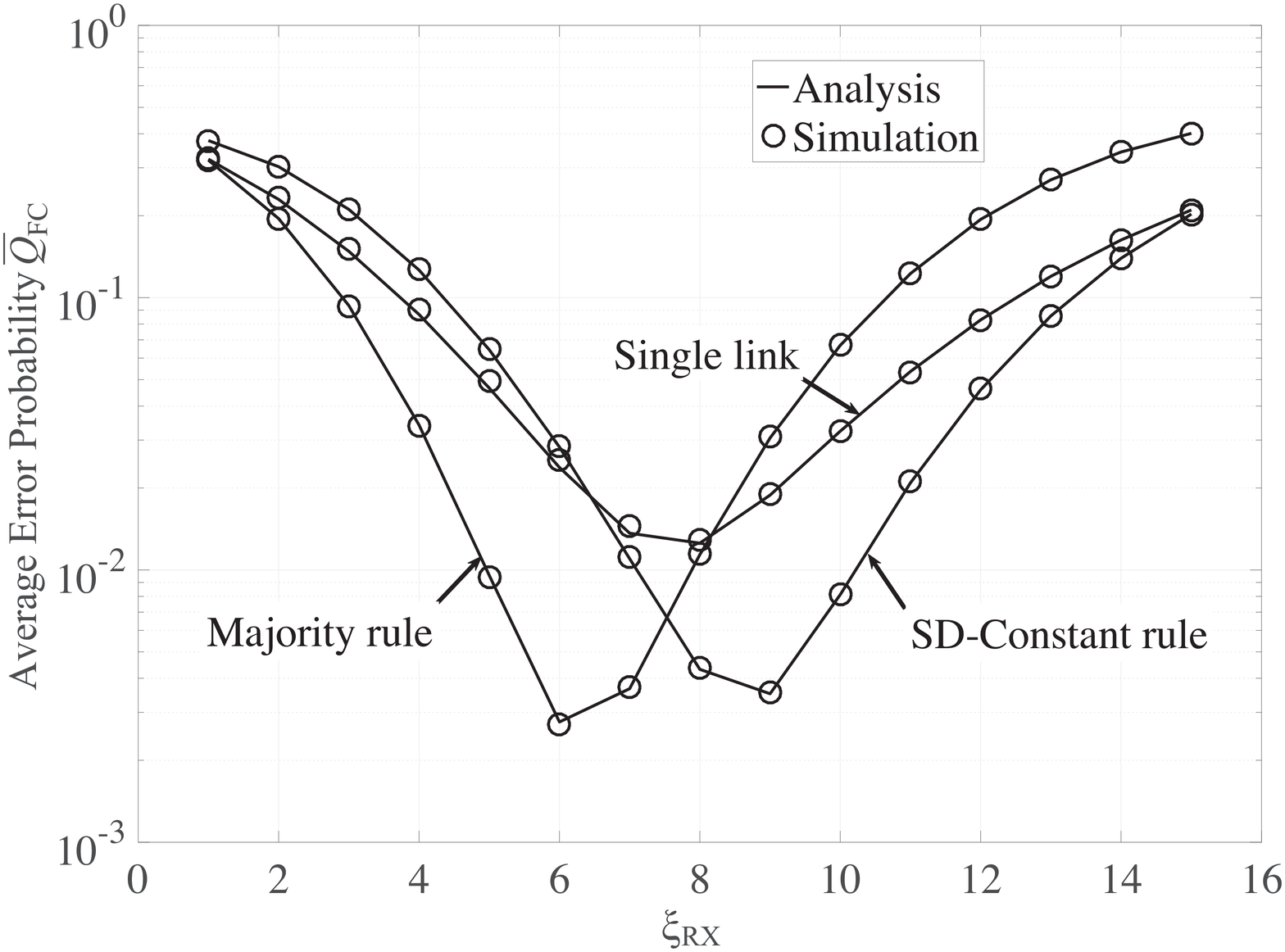}
  \caption{Average global error probability $\overline{Q}_{\scriptscriptstyle\FC}$ versus the detection threshold $\xi_{\scriptscriptstyle\R}$ with $K=3$ and $r_{\scriptscriptstyle\R}=0.225\,{\mu}\metre$ in the symmetric topology. }\label{PevsRXthreshold}
\end{figure}
\begin{figure}[t]
  \centering
  \includegraphics[height=2.2in]{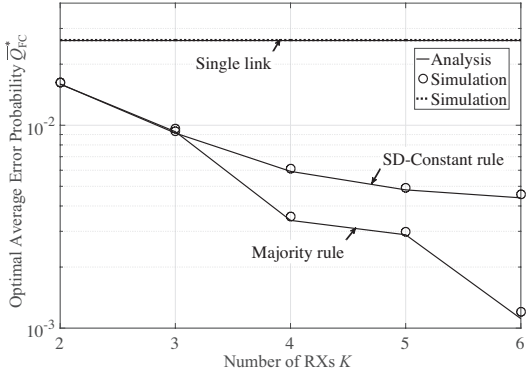}
  \caption{Optimal average global error probability $\overline{Q}_{\scriptscriptstyle\FC}^{\ast}$ versus the number of cooperative RXs, $K$, with $r_{\scriptscriptstyle\R}=0.2\,{\mu}\metre$ in the symmetric topology.}
  \label{PevsKnew}
\end{figure}
We compare the error performance of the SD-Constant scheme with those of the other two schemes. The first one is a single-link scheme, where there is a TX at ($0$, $0$, $0$) and one RX at ($2\,{\mu}\metre$, $0.6\,{\mu}\metre$, $0$) in the system (i.e., no FC). We assume that the TX releases $10000$ molecules for symbol ``1'' in the single-link scheme. The second scheme is the majority rule explored in \cite{fang2016distributed}, where the RXs report their decisions with distinct types of molecules and the FC decides using the majority rule. Except for the parameters specified above, we set the other parameters of these two schemes to be the same as those for the SD-Constant scheme.

In Fig. \ref{PevsRXthreshold}, we consider the symmetric topology and plot the average global error probability $\overline{Q}_{\scriptscriptstyle\FC}$ of a three-RX cooperative MC system versus the detection threshold at the RXs, $\xi_{\scriptscriptstyle\R}$. We consider a fixed threshold at the FC that is assumed independent of $\xi_{\scriptscriptstyle\R}$, i.e., $\xi_{\scriptscriptstyle\FC}=4$ for the majority rule and $\xi_{\scriptscriptstyle\FC}=6$ for the SD-Constant scheme. We see that the simulation curves match with the analytical curves, which demonstrates the accuracy of our analytical results. 
We also see that the SD-Constant scheme only suffers a slight $20{\%
}$ error performance degradation compared to the majority rule and the SD-Constant scheme outperforms the single link significantly by almost a factor of 3 at their corresponding optimal RX detection thresholds.

In Fig. \ref{PevsKnew}, we consider the symmetric topology and plot the optimal error probability $\overline{Q}_{\scriptscriptstyle\FC}^{\ast}$ versus the number of RXs, $K$. The value of $\overline{Q}_{\scriptscriptstyle\FC}^{\ast}$ for each $K$ is achieved by jointly numerically optimizing $\xi_{\scriptscriptstyle\R}$ and $\xi_{\scriptscriptstyle\FC}$. We see that the simulation curves agree with the analytical curves. The SD-Constant and the majority rule outperform the single link and the majority rule outperforms the SD-Constant. This is consistent with our expectation since for the majority rule, the RXs report to the FC with different types of molecules and that requires higher computational complexity.
 We further see that the error performance of the SD-Constant and the majority rules profoundly improve as $K$ increases, even though the total number of molecules used in the cooperative MC system is constrained.
\begin{figure}[t]
  \centering
  \includegraphics[height=2.2in]{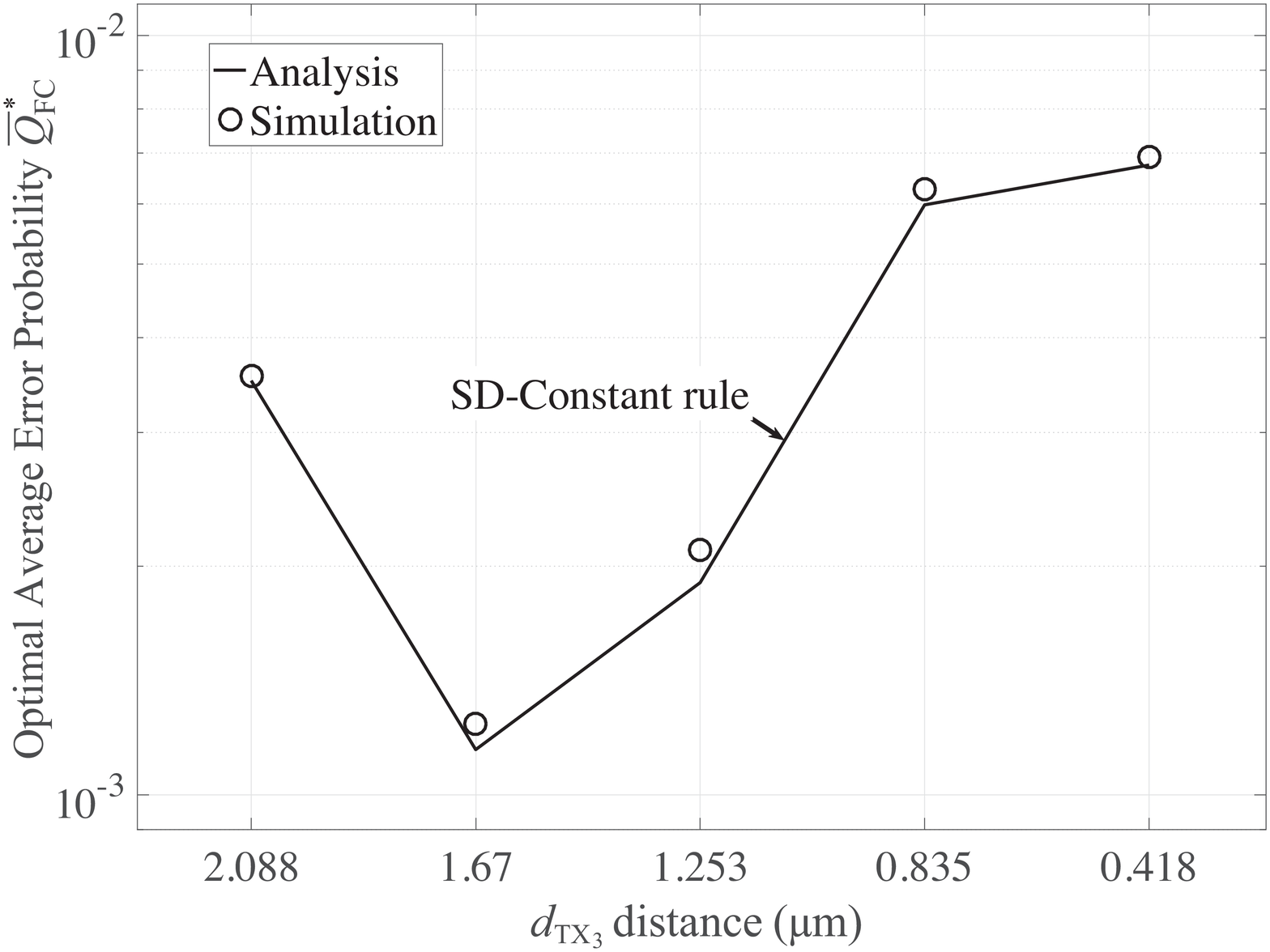}
  \caption{Optimal average global error probability $\overline{Q}_{\scriptscriptstyle\FC}^{\ast}$ versus the distance between TX and $\RX_3$, $d_{\scriptscriptstyle\TX_3}$, with $K=3$ and $r_{\scriptscriptstyle\R}=0.225\,{\mu}\metre$.}
  \label{asymmetric}
\end{figure}

In Fig. \ref{asymmetric}, we consider the asymmetric topology and plot the optimal error probability $\overline{Q}_{\scriptscriptstyle\FC}^{\ast}$ of a three-RX cooperative MC system versus the distance between the TX and $\RX_3$, $d_{\scriptscriptstyle\TX_3}$. We keep the positions of $\R_1$ and $\R_2$ fixed and move $\R_3$ along the line segment between the symmetric position and the TX, as indicated in Table \ref{tab:coordinates2}. The value of $\overline{Q}_{\scriptscriptstyle\FC}^{\ast}$ for each $d_{\scriptscriptstyle\TX_3}$ is achieved by jointly numerically optimizing $\xi_{\scriptscriptstyle\R}$ and $\xi_{\scriptscriptstyle\FC}$. We see that our analysis is again verified by simulations. The error performance first improves and then decreases as $\R_3$ moves toward the TX. We observe that $d_{\scriptscriptstyle\TX_3} = 1.67 \mu\metre$ has the best error performance. This is because both the $\TX$-$\RX_3$ link and the $\RX_3$-$\FC$ link contribute to the error performance of the system. When $d_{\scriptscriptstyle\TX_3} = 2.088 \mu\metre$, the system error performance is dominated by the $\TX$-$\RX_3$ link and this link becomes more reliable as $d_{\scriptscriptstyle\TX_3}$ decreases. For $d_{\scriptscriptstyle\TX_3}<1.67 \mu\metre$, the system error performance is dominated by the $\RX_3$-$\FC$ link, which becomes weaker when $d_{\scriptscriptstyle\TX_3}$ decreases.

\section{Conclusion}
In this paper, we proposed a practical multi-receiver cooperative scheme called SD-Constant to improve the reliability of diffusion-based MC systems. In this scheme, each RX uses the same type of molecule to report its decisions. Hence, the new scheme is more bio-realistic than hard fusion rules for environments where the types of molecules available is constrained and the devices have lower computational complexity. We also considered asymmetric and symmetric topologies for the system. We derived closed-form expressions for the global error probabilities for both topologies. Using numerical and simulation results, we verified our analysis. We demonstrated the error performance advantage of the SD-Constant scheme over the single-link approach. We note that for the SD-Constant scheme, the slight decrease in error performance is an acceptable trade-off for a computational complexity that is lower than that of the majority rule.


\end{document}